\documentclass[]{IEEEtran}

\usepackage{tikz}
\usepackage{amsmath}
\usepackage{amssymb}
\usepackage{amsfonts}
\usepackage{amsthm}
\usepackage{dsfont}
\usepackage{algorithmic}
\usepackage{algorithm}
\usepackage{graphics}
\usepackage[compress]{cite}
\usepackage{multirow}
\usepackage{url}

\newtheorem{definition}{Definition}
\newtheorem{proposition}{Proposition}
\newtheorem{theorem}{Theorem}

\DeclareMathOperator*{\argmax}{arg\,max}
\DeclareMathOperator{\sign}{sign}
\DeclareMathOperator{\Var}{Var}
\DeclareMathOperator{\erf}{erf}
\DeclareMathOperator{\trace}{trace}

\begin{document}
\title{A Game-Theoretic Approach to \\
Adversarial Linear Support Vector Classification}


\author{Farhad Farokhi
\thanks{F. Farokhi is with the CSIRO's Data61 and the University of Melbourne. E-mails: farhad.farokhi@\{data61.csiro.au,unimelb.edu.au\}}
}


\maketitle

\begin{abstract} In this paper, we employ a game-theoretic model to analyze the interaction between an adversary and a classifier. There are two classes (i.e., positive and negative classes) to which data points can belong. The adversary is interested in maximizing the probability of miss-detection for the positive class (i.e., false negative probability). The adversary however does not want to significantly modify the data point so that it still maintains favourable traits of the original class. The classifier, on the other hand, is interested in maximizing the probability of correct detection for the positive class (i.e., true positive probability) subject to a lower-bound on the probability of correct detection for the negative class (i.e., true negative probability). For conditionally Gaussian data points (conditioned on the class) and linear support vector machine classifiers, we rewrite the optimization problems of the adversary and the classifier as convex optimization problems and use best response dynamics to learn an equilibrium of the game. This results in computing a  linear support vector machine classifier that is robust against adversarial input manipulations. We illustrate the framework on a synthetic dataset and a public Cardiovascular Disease dataset.
\end{abstract}


\section{Introduction}
Rapid developments in machine learning techniques is anticipated to boost productivity and spur economic growth. The potential to extract accurate analytic gives rise to a data-driven economy which, according to a recent McKinsey report~\cite{McKinseyreport}, is estimated to potentially deliver an additional economic output of around \$13 trillion by 2030. This has motivated a world-wide race to put machine learning in everything ranging from health sector to aerospace engineering. However, machine learning systems face security concerns that, up to recently, have not attracted much attention. 

Machine learning algorithms have been observed to be vulnerable to adversarial manipulations of their inputs after training and deployment, known as evasion attacks~\cite{dalvi2004adversarial, goodfellow2014explaining,yuan2019adversarial}. In fact, some machine learning models are shown to be adversely influenced by very small perturbations to the inputs~\cite{biggio2013evasion, goodfellow2014explaining, papernot2016distillation}. These observations severely restrict their applications in practice. 

Most common methods for securing machine learning algorithms against adversarial inputs are \textit{ad hoc} in nature or based on heuristic; see, e.g.,~\cite{papernot2016distillation, goodfellow2014explaining,kurakin2016adversarial}. For instance, it has been shown that injecting adversarial examples into the training set, often referred to as adversarial training, can increase robustness to adversarial manipulations~\cite{goodfellow2014explaining}. However, this approach is dependant on the method used for generating adversarial examples and the number of the required adversarial examples is often not known \textit{a priori}. 

In this paper, we propose a game-theoretic approach to model and analyze the interactions between an adversary and a decision maker (i.e., a classifier). As a starting point for research, we focus on a binary classification problem using linear support vector machines with Gaussian-distributed data in each class. This way, we can compute optimal adversarial linear support vector machine. Note that the problem of detecting and mitigating evasion attacks in support vector machines is still an ongoing debate~\cite{frederickson2018attack, han2018adequacy} with not much known in the way of designing robust classifiers, except through adding adversarial examples to the training dataset (discussed in the earlier references) or, in a heuristical manner, by changing the regularization term~\cite{russu2016secure}.

We particularly model the interaction between the adversary and the classifier using a constant-sum game. There are two classes (i.e, positive and negative classes) to which the data can belong. The adversary is interested in maximizing the probability of miss-detection for the positive class, i.e., the probability of classification of an input belonging to the negative class while it is from the positive classes, also known as the false negative probability. However, the adversary does not want to significantly modify the data so that it still maintains the favourable traits of the original class. An example of such a classification problem is a simplified spam filtering in which the nature of the email determines its class (with the  positive class denoting spam emails). The adversary's objective is to modify the spam emails so that they pass the spam filtering algorithm. Manipulating the email by a large amount might negate the adversarial nature of spam emails. Also, note that the adversary cannot access all the emails and thus can only manipulate the spam emails.  The classifier is interested in maximizing the probability of correct detection for the positive class, i.e., the probability of classification of an input belonging to the positive class if it is from the positive classes, also known as the true positive probability. In the spam filtering example, the classifier aims to determine if an email is spam or not based on possibly modified spam email and unaltered genuine emails. Evidently, if the objective of the classifier was solely to correctly catch all data points belonging to the positive class, its optimal behaviour would have been to ignore the received data point and to mark it as belonging to the positive class. This would correctly identify all data points belonging to the positive classes however it also miss-classifies all data points from the negative class. This is in fact impractical. For instance, such a policy, in the spam filtering example, would results in marking all emails as spam, which is undesirable. Therefore, the classifier enforces a lower bound on the probability of correct detection for the negative class, i.e., the probability of classification of an input belonging to the negative class if it is from the negative class, also known as the true negative probability. We rewrite the optimization problems of the adversary and the classifier as two convex optimization problems and use a best response dynamics to learn an equilibrium of the game. 

The problem formulation of this paper is in essence close to cheap-talk games~\cite{crawford1982strategic,farokhi2016estimation,farrell1996cheap} and Bayesian persuasion~\cite{kamenica2011bayesian,dughmi2016algorithmic, nadendla2018effects} in which a better-informed sender wants to communicate with a receiver in a strategic manner to sway its decision. However, there is a stark difference between those studies and the setup of this paper. In this paper, the classifier (i.e., the receiver) is restricted to follow a machine learning model (specifically, a linear support vector machine), which is not necessarily Bayesian. 

The rest of the paper is organized as follows. Section~\ref{sec:problem} presents the game-theoretic problem formulation. Numerical methods for computing the equilibrium are presented in Section~\ref{sec:results}. Numerical examples are presented in Section~\ref{sec:numerical}. Finally, Section~\ref{sec:conclusions} concludes the paper and presents avenues for future research. 

\section{Problem Formulation} \label{sec:problem}

\begin{figure}
    \centering
    \begin{tikzpicture}
    \node[draw,rectangle,minimum width=2cm,minimum height=1cm] (S) at (0,0.0) {Adversary};
    \node[draw,rectangle,minimum width=2cm,minimum height=1cm] (R) at (3,0.0) {Classifier};
    \draw[->] (S) -- node[above] {$y$} (R);
    \draw[->] (-2,0) -- node[above] {$x$} (S);
    \draw[->] (R) -- node[above] {$z$} (5,0);
    \end{tikzpicture}
    \caption{Communication structure between an adversary and a classifier playing the adversarial classification game.}
    \label{fig:structure}
\end{figure}
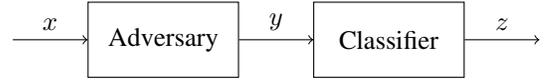

Consider the communication structure between an adversary and a classifier as in Figure~\ref{fig:structure}. 

The \textit{adversary} has access to a random variable $x\in\mathbb{R}^n$, which can belong to two classes: positive and negative. The class to which $x$ belongs is denoted by $\theta\in\{-1,+1\}$, which is  a binary random variable itself with $\mathbb{P}\{\theta=+1\}=1-\mathbb{P}\{\theta=-1\}=\alpha>0$.
The random variable $x$ is assumed to be Gaussian with mean $\mu_+\in\mathbb{R}^{n}$ and co-variance matrix $\Sigma_+\succ 0$ if $\theta=+1$ and is assumed to be Gaussian with mean $\mu_-\in\mathbb{R}^{n}$ and co-variance matrix $\Sigma_-\succ 0$ if $\theta=-1$. The notation $A\succ 0$ implies that $A$ is a symmetric positive definite matrix while $A\succeq 0$ implies that $A$ is positive semi-definite.  

The adversary communicates a message $y\in\mathbb{R}^n$ to the classifier. This message may or may not be truthful. We assume that $y$ follows 
\begin{align} \label{eqn:linear_adversary}
    y=
    \begin{cases}
    Ax+w, & \theta=+1,\\
    x, & \theta=-1,
    \end{cases}
\end{align}
where $A\in\mathbb{R}^{n\times n}$ is a weighting matrix and $w\in\mathbb{R}^m$ is a Gaussian random variable (i.e., additive Gaussian noise) with mean $\mu_w\in\mathbb{R}^m$ and co-variance $\Sigma_w\succeq 0$. Let $\gamma:=(A,\mu_w,\Sigma_w)$ be the \textit{policy of the adversary}. The set of all policies of the adversary is denoted by $\Gamma$. Note that the adversary may not be able to access all the data points, e.g., all the emails in the spam filtering example discussed in the introduction, and thus can only manipulate the ones belonging to the positive class, e.g., spam emails. 

In this paper, we restrict the adversary's policy to be linear. This is to ensure that the adversary's abilities is a match for the classifier (as the classifier is assumed to be linear support vector machine). Furthermore, the linearity of the adversary simplifies the analysis due to the Gaussianity of the data points. This analysis provides a lower bound
on the influence of a more general adversary  since they can find more degrees of freedom for manipulating the data points by extending their set of strategies to also cover nonlinear mappings.

The \textit{classifier} intends to determine the class to which the random variable $x$ actually belongs,  based on the received message $y$. The decision of the classifier is denoted by $z\in\{-1,+1\}$ and is determined by a linear support vector machine classifier:
\begin{align} \label{eqn:linear_svm}
    z=\sign(\alpha^\top y+\beta),
\end{align}
where $\alpha\in\mathbb{R}^n$ is a vector of weights and $\beta\in\mathbb{R}$ is a bias. Note that scaling both $\alpha$ and $\beta$ by a positive constant does not change the sign of $\alpha^\top y+\beta$ and thus, without loss of generality, we can assume that $\|\alpha\|_\infty\leq 1$ and $|\beta|\leq 1$. Let $\eta:=(\alpha,\beta)$ be the \textit{policy of the classifier}. The set of all policies of the classifier is denoted by $\Upsilon$.


The goal of the classifier is to correctly classify as many cases  belonging to the positive class  as possible. Therefore, the classifier wants to maximize 
\begin{align}
    U_{\mathrm{c}}(\gamma,\eta):=\mathbb{P}\{z=+1|\theta=+1\}.
\end{align}
Evidently, if the objective of the classifier was solely to maximize $\mathbb{P}\{z=+1|\theta=+1\}$, the optimal behaviour would have been to set $\beta=1$ and $\alpha=0$, i.e., ignore the received messages and mark it as belonging to the positive class. This is however impractical, e.g., such a policy in the spam filtering example results in marking all emails as spam which is undesirable. Therefore, the set of the actions of the classifier is constrained by 
\begin{align}
    C_{\mathrm{c}}(\gamma,\eta):=\mathbb{P}\{z=-1|\theta=-1\}\geq 1-\delta,
\end{align}
where $\delta\in(0,1/2)$ is a constant. By selecting $\delta$ small enough, we can ensure that data points belonging to the negative class (e.g., non-spam emails) are almost entirely correctly classified. In what follows, we use the notation \begin{align}
    \Omega_{\mathrm{c}}(\gamma):=\{\eta=(\alpha,\beta)\in\Upsilon:&C_{\mathrm{c}}(\gamma,\eta)\geq 1-\delta,\nonumber\\
    &\|\alpha\|_\infty\leq 1,|\beta|<1\}.
\end{align}

The goal of the adversary is to deceive the classifier into miss-classifying more data points from the positive class, e.g., accepting more spam emails. This is achieved by maximizing 
\begin{align}
    U_{\mathrm{a}}(\gamma,\eta):=\mathbb{P}\{z=-1|\theta=+1\}.
\end{align}
However, the adversary does not want to make large changes to $x$ as that might defeat its original purpose, e.g., in the spam filtering example, the email might no longer contain the desirable traits of the spam emails like unsolicited commercial advertisements. Conceptually, this can be achieved by ensuring that the magnitude of the changes is constrained by 
\begin{align}
    C_{\mathrm{a}}(\gamma,\eta):=\mathbb{E}\{\|x-y\|_2^2|\theta=+1\}\leq \epsilon,
\end{align}
where $\epsilon>0$ is a constant. Although, in this paper, we consider a constraint on the variance of the manipulations, the analysis can be readily extended to other constraints, e.g., mean absolute deviations.  In what follows, we use the notation 
\begin{align}
\Omega_{\mathrm{a}}(\eta):= \{\gamma\in\Gamma:C_{\mathrm{a}}(\gamma,\eta)\leq \epsilon\}.
\end{align}

\begin{definition}[Adversarial Classification Game] An adversarial classification game is defined as a strategic game between two players: an adversary and a classifier. The utilities of the adversary and the classifier are $U_{\mathrm{c}}(\gamma,\eta)$ and $U_{\mathrm{a}}(\gamma,\eta)$, respectively. The action spaces of the adversary and the classifier are $\Omega_{\mathrm{c}}(\gamma,\eta)$ and $\Omega_{\mathrm{a}}(\gamma,\eta)$, respectively.
\end{definition}

In the language of~\cite{arrow1954existence}, an adversarial classification game is in fact a competitive economy as the action spaces of the players potentially depends on the actions of the other players. In this paper, we use game instead of competitive economy in line with more recent game theory literature.

\begin{definition}[Equilibrium] A pair of policies $(\gamma^*,\eta^*)$ constitutes an equilibrium if
\begin{subequations}
\begin{align}
    \gamma^*\in&\argmax_{\gamma\in\Omega_{\mathrm{a}}(\eta^*)} U_{\mathrm{a}}(\gamma,\eta^*),
    \label{eqn:adversary_optimization}\\
    \eta^*\in&\argmax_{\eta\in \Omega_{\mathrm{c}}(\gamma^*)}  U_{\mathrm{c}}(\gamma^*,\eta).
    \label{eqn:classifier_optimization}
\end{align}
\end{subequations}
\end{definition}
With these definitions in hand, we are ready to present the results of the paper.

\section{Main Results} \label{sec:results}

We can prove an important result regarding the adversarial classification game illustrating the direct conflict of interest between the adversary and the classifier, as expected.

\begin{proposition}[Constant-Sum Game] \label{prop:1} The adversarial classification game is a constant-sum game.
\end{proposition}

\begin{IEEEproof} Note that $U_{\mathrm{c}}(\gamma,\eta) +U_{\mathrm{a}}(\gamma,\eta)=\mathbb{P}\{z=+1|\theta=+1\}+\mathbb{P}\{z=-1|\theta=+1\}=1$ for any $\gamma$ and $\eta$.
\end{IEEEproof}

In the the remainder of this section, we provide a method for computing equilibria of an adversarial classification game. We first show that the best responses of the players can be computed using convex optimizations. In what follows, $\erf:\mathbb{R}\rightarrow[0,1]$ denotes the error function, defined as $ \erf(x):=(\int_{-x}^x \exp(-t^2)\mathrm{d}t)/\sqrt{\pi}.$

\begin{theorem} Let $(\gamma^*,\eta^*)$ be such that~\eqref{eqn:linear_adversary} and~\eqref{eqn:linear_svm} holds with probability one with the following parameters:
\begin{align*}
    A&=\overline{A}/t_1, \mu_w=\overline{\mu}_w/t_1, \Sigma_w=R_wR_w^\top/t_1,\\
    \alpha&=\overline{\alpha}/\max\{\|\overline{\alpha}\|_\infty,|\overline{\beta}|\},  \beta=\overline{\beta}/\max\{\|\overline{\alpha}\|_\infty,|\overline{\beta}|\},
\end{align*}
where $(\overline{A},\overline{\mu}_w,R_w,Z',t_1)$ is given by
\begin{subequations}\label{eqn:adv_best_response}
\begin{align} 
\argmax_{\overline{A},\overline{\mu}_w,R_w,Z',t} &\quad -\alpha^\top (\overline{A}\mu_++\overline{\mu}_w)-\beta\\
    \mathrm{s.t.}\quad\;\;\, & \quad 
    \begin{bmatrix}
    \frac{1}{\sqrt{2}} I & 0 & R_w^\top \alpha \\
    0 & \Sigma_+^{-1} & \overline{A}^\top\alpha \\
    \alpha^\top R_w & \alpha^\top\overline{A} & t
    \end{bmatrix}
    \succeq 0
    ,\\
    &\quad  \trace(Z')\leq t\epsilon,\\
    &\quad 
    \begin{bmatrix}
    tI & 0 & 0 & R_w^\top \\
    0 & t & 0 & \overline{\mu}_w^\top \\
    0 & 0 & t(\Sigma_+\hspace*{-.03in}+\hspace*{-.03in}\mu_+\mu_+^\top )^{\hspace*{-.01in}-\hspace*{-.01in}1} & tI-\overline{A}^\top \\
    R_w & \overline{\mu}_w & tI-\overline{A} & Z'
    \end{bmatrix}\hspace*{-.05in}\succeq \hspace*{-.03in}0, \\
    &\quad t\geq 0.
\end{align}
\end{subequations}
and $ (\overline{\alpha},\overline{\beta},t_2)$ is given by
\begin{subequations}\label{eqn:classifier_best_response}
\begin{align}
\argmax_{\overline{\alpha},\overline{\beta},t} \quad &\overline{\alpha}^\top (A\mu_++\mu_w)+\overline{\beta}, \\
    \mathrm{s.t.}\quad\quad & 
    2\overline{\alpha}^\top (A\Sigma_+A^\top+\Sigma_w) \overline{\alpha}\leq 1,\\
    & -\overline{\alpha}^\top\mu_--\overline{\beta}  \geq\erf^{-1}\left( 1-2\delta\right) t,\\
    & \left\|\sqrt{2}\Sigma_-^{1/2}\overline{\alpha}\right\|_2\leq t,\\
    & t\geq 0.
\end{align}
\end{subequations}
Then $(\gamma^*,\eta^*)$ is an equilibrium of the adversarial classification game.
\end{theorem}

\begin{IEEEproof} Let us consider the classifier. The utility of the classifier can be simplified as
\begin{align*}
    U_{\mathrm{c}}(\gamma,\eta)
    =&\mathbb{P}\{z=+1|\theta=+1\}\\
    =&\mathbb{P}\{\alpha^\top y+\beta>0|\theta=+1\}\\
    =&\frac{1}{2}-\frac{1}{2}\erf\left(\frac{-\alpha^\top (A\mu_++\mu_w)-\beta}{\sqrt{2\alpha^\top (A\Sigma_+A^\top+\Sigma_w) \alpha}} \right),
\end{align*}
where the last equality follows from that $\alpha^\top y+\beta$, conditioned on the observation that $\theta=+1$, is a Gaussian random variable with the following mean and variance:
\begin{align*}
    \mathbb{E}\{\alpha^\top y+\beta|\theta=+1\}&=\alpha^\top (A\mu_++\mu_w)+\beta,\\
    \Var(\alpha^\top y+\beta|\theta=+1)&=\alpha^\top (A\Sigma_+A^\top +\Sigma_w) \alpha .
\end{align*}
Because $\erf(\cdot)$ is an increasing function, maximizing $U_{\mathrm{c}}(\gamma,\eta)$ is equivalent to minimizing $(-\alpha^\top (A\mu_++\mu_w)-\beta)/(\sqrt{2\alpha^\top (A\Sigma_+A^\top+\Sigma_w) \alpha})$. The constraint of the classifier can also be rewritten as
\begin{align*}
    C_{\mathrm{c}}(\gamma,\eta)
    =&\mathbb{P}\{z=-1|\theta=-1\}\\
    =&\mathbb{P}\{\alpha^\top y+\beta<0|\theta=-1\}\\
    =&\frac{1}{2}+\frac{1}{2}\erf\left(\frac{-\alpha^\top\mu_--\beta}{\sqrt{2\alpha^\top \Sigma_- \alpha}} \right)
\end{align*}
where the last equality again follows from that $\alpha^\top y+\beta$, conditioned on the observation that $\theta=-1$, is a Gaussian random variable with the following mean and variance:
\begin{align*}
    \mathbb{E}\{\alpha^\top y+\beta|\theta=-1\}&=\alpha^\top\mu_-+\beta,\\
    \Var(\alpha^\top y+\beta|\theta=-1)&=\alpha^\top \Sigma_- \alpha .
\end{align*}
The constraint that $C_{\mathrm{c}}(\gamma,\eta)\geq 1-\delta$ is equivalent to 
\begin{align*}
    \erf\left(\frac{-\alpha^\top\mu_--\beta}{\sqrt{2\alpha^\top \Sigma_- \alpha}} \right)\geq 1-2\delta.
\end{align*}
Again, because $\erf(\cdot)$ is an increasing function, we can rewrite the classifier's constraint as
\begin{align*}
     \frac{-\alpha^\top\mu_--\beta}{\sqrt{2\alpha^\top \Sigma_- \alpha}}  \geq\delta':=\erf^{-1}\left( 1-2\delta\right).
\end{align*}
Note that $\erf^{-1}\left( 1-2\delta\right)>0$ because $\delta\in(0,1/2)$. 
These derivations allow us to transform the optimization problem in~\eqref{eqn:classifier_optimization} into
\begin{subequations} \label{eqn:classifier_3}
\begin{align}
    \max_{\alpha,\beta} \quad &\frac{\alpha^\top (A\mu_++\mu_w)+\beta}{\sqrt{2\alpha^\top (A\Sigma_+A^\top+\Sigma_w) \alpha}}, \\
    \mathrm{s.t.}\quad & \frac{-\alpha^\top\mu_--\beta}{\sqrt{2\alpha^\top \Sigma_- \alpha}}  \geq\delta',\\
    & \|\alpha\|_\infty \leq 1,\\
    & |\beta|\leq 1.
\end{align}
\end{subequations}
We use the approach of~\cite{schaible1974parameter} for the constraint to eliminate the fractional constraint. We define the change of variables:
\begin{align*}
    g&=\frac{1}{\sqrt{2\alpha^\top \Sigma_- \alpha}},\;
    \tilde{\alpha}=\frac{1}{\sqrt{2\alpha^\top \Sigma_- \alpha}} \alpha,\;
    \tilde{\beta}=\frac{1}{\sqrt{2\alpha^\top \Sigma_- \alpha}} \beta.
\end{align*}
Hence, we can rewrite~\eqref{eqn:classifier_3} as 
\begin{subequations}
\begin{align}
    \max_{\alpha,\beta} \quad &\frac{\tilde{\alpha}^\top (A\mu_++\mu_w)+\tilde{\beta}}{\sqrt{2\tilde{\alpha}^\top (A\Sigma_+A^\top+\Sigma_w) \tilde{\alpha}}}, \\
    \mathrm{s.t.}\quad & -\tilde{\alpha}^\top\mu_--\tilde{\beta}\geq \delta'\\
                    & 2\tilde{\alpha}^\top \Sigma_-\tilde{\alpha}\leq 1,\\
                    &\|\tilde{\alpha}\|_\infty \leq g,\\
                    & |\tilde{\beta}|\leq g.
\end{align}
\end{subequations}
We can drop the last two inequalities noting that we can set $g=\max\{\|\alpha^*\|_\infty,|\beta^*|\}$, where 
\begin{subequations} \label{eqn:classifier_5}
\begin{align}
    (\alpha^*,\beta^*)\in \argmax_{\alpha,\beta} \quad &\frac{\tilde{\alpha}^\top (A\mu_++\mu_w)+\tilde{\beta}}{\sqrt{2\tilde{\alpha}^\top (A\Sigma_+A^\top+\Sigma_w) \tilde{\alpha}}}, \\
    \mathrm{s.t.}\quad\quad  & -\tilde{\alpha}^\top\mu_--\tilde{\beta}\geq \delta',\\
    & 2\tilde{\alpha}^\top \Sigma_-\tilde{\alpha}\leq 1.
\end{align}
\end{subequations}
Again, we use the approach of~\cite{schaible1974parameter}, but this time for the utility, to rewrite this fractional optimization problem. To do so, define 
\begin{align*}
    t&=\frac{1}{\sqrt{2\tilde{\alpha}^\top (A\Sigma_+A^\top+\Sigma_w) \tilde{\alpha}}}\\
    \overline{\alpha}&=\frac{1}{\sqrt{2\tilde{\alpha}^\top (A\Sigma_+A^\top+\Sigma_w) \tilde{\alpha}}} \tilde{\alpha},\\
    \overline{\beta}&=\frac{1}{\sqrt{2\tilde{\alpha}^\top (A\Sigma_+A^\top+\Sigma_w) \tilde{\alpha}}}\tilde{\beta},
\end{align*}
Following~\cite{schaible1974parameter}, the optimization problem in~\eqref{eqn:classifier_5} is equivalent with
\begin{subequations}
\begin{align}
    \max_{\overline{\alpha},\overline{\beta},t} \quad &\overline{\alpha}^\top (A\mu_++\mu_w)+\overline{\beta}, \\
    \mathrm{s.t.}\,\quad & 
    2\overline{\alpha}^\top (A\Sigma_+A^\top+\Sigma_w) \overline{\alpha}\leq 1,\\
    & -\overline{\alpha}^\top\mu_--\overline{\beta}  \geq\delta' t,\\
    & \left\|\sqrt{2}\Sigma_-^{1/2}\overline{\alpha}\right\|_2\leq t,\\
    & t\geq 0.
\end{align}
\end{subequations}
Now, we consider the adversary. The utility of the adversary is given by
\begin{align*}
    U_{\mathrm{a}}(\gamma,\eta)
    =&\mathbb{P}\{z=-1|\theta=+1\}\\
    =&1-\mathbb{P}\{z=+1|\theta=+1\}\\
    =&1-U_{\mathrm{c}}(\gamma,\eta)\\
    =&\frac{1}{2}+\frac{1}{2}\erf\left(\frac{-\alpha^\top (A\mu_++\mu_w)-\beta}{\sqrt{2\alpha^\top (A\Sigma_+A^\top+\Sigma_w) \alpha}} \right)
\end{align*}
Hence, because $\erf(\cdot)$ is an increasing function, maximizing $U_{\mathrm{a}}(\gamma,\eta)$ is the same as maximizing
$(-\alpha^\top (A\mu_++\mu_w)-\beta)/\sqrt{2\alpha^\top (A\Sigma_+A^\top+\Sigma_w) \alpha}$. 
Furthermore, the constraint of the adversary can be rewritten as 
\begin{align*}
    C_{\mathrm{a}}(\gamma,\eta)
    =&\mathbb{E}\{\|x-y\|_2^2|\theta=+1\}\\
    =&\mathbb{E}\{\|(I-A)x-w\|_2^2|\theta=+1\}\\
    =&\trace((I-A)(\Sigma_++\mu_+\mu_+^\top )(I-A)^\top)\\
    &+\trace(\mu_w\mu_w^\top +\Sigma_w)
\end{align*}
Following these derivations, we can transform the optimization problem in~\eqref{eqn:adversary_optimization} to 
\begin{subequations} \label{eqn:adversary_2}
\begin{align}
    \max_{A,\Sigma_w,\mu_w} & \quad \frac{-\alpha^\top (A\mu_++\mu_w)-\beta}{\sqrt{2\alpha^\top (A\Sigma_+A^\top+\Sigma_w) \alpha}}\\
    \mathrm{s.t.}\quad &\quad  \trace((I-A)(\Sigma_++\mu_+\mu_+^\top )(I-A)^\top)\nonumber\\
    &\hspace{.6in}+\trace(\mu_w\mu_w^\top +\Sigma_w)\leq \epsilon,\label{eqn:const1}\\
    &\quad \Sigma_w\succeq 0.
\end{align}
\end{subequations}
Note that the inequality constraint~\eqref{eqn:const1} can be replaced with the following three constraints:
\begin{subequations}
\begin{align}
    (I-A)(\Sigma_++\mu_+\mu_+^\top )(I-A)^\top +\mu_w\mu_w^\top+\Sigma_w\preceq& Z,\label{eqn:const2}\\
    Z\succeq& 0,\\
    \trace(Z)\leq &\epsilon.
\end{align}
\end{subequations}
Further, the constraint~\eqref{eqn:const2} can be linearized using its Schur complement because $\Sigma_+\succ 0$ to get
\begin{align}\label{eqn:const2.5}
\begin{bmatrix}
(\Sigma_++\mu_+\mu_+^\top )^{-1} & I-A^\top \\
I-A & Z-\Sigma_w-\mu_w\mu_w^\top
\end{bmatrix}\succeq 0.
\end{align}
Again, we can use the Schur complement, to transform the constraint~\eqref{eqn:const2.5} into
\begin{align}
\begin{bmatrix}
1 & 0 & \mu_w^\top \\
0 & (\Sigma_++\mu_+\mu_+^\top )^{-1} & I-A^\top \\
\mu_w & I-A & Z-\Sigma_w
\end{bmatrix}\succeq 0.
\end{align}
Therefore, the optimization problem in~\eqref{eqn:adversary_2} can be transformed into
\begin{subequations} \label{eqn:adversary_3}
\begin{align}
    \max_{A,\Sigma_w,\mu_w,Z} \quad  & \frac{-\alpha^\top (A\mu_++\mu_w)-\beta}{\sqrt{2\alpha^\top (A\Sigma_+A^\top+\Sigma_w) \alpha}}\\
    \mathrm{s.t.}\quad \;\quad&  \begin{bmatrix}
1 & 0 & \mu_w^\top \\
0 & (\Sigma_++\mu_+\mu_+^\top )^{-1} & I-A^\top \\
\mu_w & I-A & Z-\Sigma_w
\end{bmatrix}\succeq 0,\\
    & Z\succeq 0,\\
    &\trace(Z)\leq \epsilon,\\
    & \Sigma_w\succeq 0.
\end{align}
\end{subequations}
Define 
\begin{align*}
    t&=\frac{1}{\sqrt{2\alpha^\top (A\Sigma_+A^\top+\Sigma_w) \alpha}},\\
    \overline{A}&=\frac{1}{\sqrt{2\alpha^\top (A\Sigma_+A^\top+\Sigma_w) \alpha}}A,\\
    \overline{\mu}_w&=\frac{1}{\sqrt{2\alpha^\top (A\Sigma_+A^\top+\Sigma_w) \alpha}}\mu_w,\\
    \overline{\Sigma}_w&=\frac{1}{\sqrt{2\alpha^\top (A\Sigma_+A^\top+\Sigma_w) \alpha}}\Sigma_w.
\end{align*}
Following~\cite{schaible1974parameter}, the optimization problem in~\eqref{eqn:adversary_3} is equivalent with
\begin{subequations} \label{eqn:adversary_4}
\begin{align}
    \max_{\overline{A},\overline{\Sigma}_w,\overline{\mu}_w,Z,t} & \quad -\alpha^\top (\overline{A}\mu_++\overline{\mu}_w)-\beta\\
    \mathrm{s.t.}\quad \;\;& \quad \alpha^\top (\overline{A}\Sigma_+\overline{A}^\top+t\overline{\Sigma}_w) \alpha\sqrt{2}\leq t,\label{eqn:const3}\\
    &\quad 
    \begin{bmatrix}
    t & 0 & \overline{\mu}_w^\top \\
    0 & t(\Sigma_++\mu_+\mu_+^\top )^{-1} & tI-\overline{A}^\top \\
    \overline{\mu}_w & tI-\overline{A} & tZ-\overline{\Sigma}_w
    \end{bmatrix}\succeq 0,\\
    &\quad  \trace(Z)\leq \epsilon,\\
    &\quad \overline{\Sigma}_w\succeq 0, Z\succeq 0, t\geq 0.
\end{align}
\end{subequations}
\begin{algorithm}[t]
\caption{\label{alg:1} Learning an equilibrium of the adversarial classification game.}
\begin{algorithmic}[1]
\REQUIRE $\varpi\in(0,1)$
\ENSURE $\gamma^{(k)}$, $\eta^{(k)}$
\FOR{$k=1,2,\dots$}
\STATE Compute  $(\overline{A},\overline{\mu}_w,R_w,Z',t_1)$ by solving~\eqref{eqn:adv_best_response} for fixed $(\alpha^{(k)\top},\beta^{(k)})$ 
\STATE Update
\begin{align*}
    A^{(k+1)}&\leftarrow\left(1-\frac{\varpi}{k}\right)A^{(k)}+\frac{\varpi}{k}\frac{1}{t_1}\overline{A},\\
    \mu_w^{(k+1)}&\leftarrow\left(1-\frac{\varpi}{k}\right)\mu_w^{(k)}+\frac{\varpi}{k}\frac{1}{t_1}\overline{\mu}_w,\\
    \Sigma_w^{(k+1)}&\leftarrow\left(1-\frac{\varpi}{k}\right)\Sigma_w^{(k)}+\frac{\varpi}{k}\frac{1}{t_1}R_w^\top R_w.
\end{align*}
\STATE Compute $(\overline{\alpha},\overline{\beta},t_2)$ by solving~\eqref{eqn:classifier_best_response} for fixed $(A^{(k)},\mu_w^{(k)},\Sigma_w^{(k)})$
\STATE Update
\begin{align*}
    \alpha^{(k+1)}&\leftarrow\left(1-\frac{\varpi}{k}\right)\alpha^{(k)}+\frac{\varpi}{k}\frac{1}{\max\{\|\overline{\alpha}\|_\infty,|\overline{\beta}|\}}\overline{\alpha},\\
    \beta^{(k+1)}&\leftarrow\left(1-\frac{\varpi}{k}\right)\beta^{(k)}+\frac{\varpi}{k}\frac{1}{\max\{\|\overline{\alpha}\|_\infty,|\overline{\beta}|\}}\overline{\beta}.
\end{align*}
\ENDFOR
\end{algorithmic}
\end{algorithm}
Using the Schur complement of~\eqref{eqn:const3} and defining $Z'=tZ$, we can transform~\eqref{eqn:adversary_4} into  
\begin{subequations}\label{eqn:adversary_5}
\begin{align} 
    \max_{\overline{A},\overline{\Sigma}_w,\overline{\mu}_w,Z',t} & \quad -\alpha^\top (\overline{A}\mu_++\overline{\mu}_w)-\beta\\
    \mathrm{s.t.}\quad\;\; & \quad 
    \begin{bmatrix}
    \Sigma_+^{-1} & \overline{A}^\top\alpha \\
    \alpha^\top\overline{A} & t-t\alpha^\top\overline{\Sigma}_w\alpha \sqrt{2}
    \end{bmatrix}
    \succeq 0\label{eqn:const4}
    ,\\
    &\quad  \trace(Z')\leq t\epsilon,\\
    &\quad 
    \begin{bmatrix}
    t & 0 & \overline{\mu}_w^\top \\
    0 & t(\Sigma_++\mu_+\mu_+^\top )^{-1} & tI-\overline{A}^\top \\
    \overline{\mu}_w & tI-\overline{A} & Z'-\overline{\Sigma}_w
    \end{bmatrix}\succeq 0,\label{eqn:const5}\\
    &\quad \overline{\Sigma}_w\succeq 0,\\
    &\quad t\geq 0.
\end{align}
\end{subequations}
By defining $R_w$ such that $R_wR_w^\top=t\overline{\Sigma}_w$ and using the Schur complements of~\eqref{eqn:const4} and~\eqref{eqn:const5}, we can transform~\eqref{eqn:adversary_5} into   
\begin{align*}
    \max_{\overline{A},R_w,\overline{\mu}_w,Z',t} & \quad -\alpha^\top (\overline{A}\mu_++\overline{\mu}_w)-\beta\\
    \mathrm{s.t.}\quad\;\;\, & \quad 
    \begin{bmatrix}
    \frac{1}{\sqrt{2}} I & 0 & R_w^\top \alpha \\
    0 & \Sigma_+^{-1} & \overline{A}^\top\alpha \\
    \alpha^\top R_w & \alpha^\top\overline{A} & t
    \end{bmatrix}
    \succeq 0
    ,\\
    &\quad  \trace(Z')\leq t\epsilon,\\
    &\quad 
    \begin{bmatrix}
    tI & 0 & 0 & R_w^\top \\
    0 & t & 0 & \overline{\mu}_w^\top \\
    0 & 0 & t(\Sigma_+\hspace*{-.03in}+\hspace*{-.03in}\mu_+\mu_+^\top )^{\hspace*{-.01in}-\hspace*{-.01in}1} & tI-\overline{A}^\top \\
    R_w & \overline{\mu}_w & tI-\overline{A} & Z'
    \end{bmatrix}\hspace*{-.05in}\succeq \hspace*{-.03in}0, \\
    &\quad t\geq 0.
\end{align*}
This concludes the proof. 
\end{IEEEproof}

We can use the best response dynamics, summarized in Algorithm~\ref{alg:1}, to extract an equilibrium of the game. The iterates in Algorithm~\ref{alg:1} converge to an equilibrium of the adversarial classification game. The convergence of the best response dynamics follows from~\cite{10230740590698} because of the constant-sum nature of the game; see Proposition~\ref{prop:1}.

\begin{figure*}[t]
\footnotesize
 \begin{tabular}{ccc}
 $\mathbb{P}\{z=-1|\theta=-1\}=0.9900$
 &
 $\mathbb{P}\{z=-1|\theta=-1\}=0.9900$
 &
 $\mathbb{P}\{z=-1|\theta=-1\}=0.9900$
 \\
 $\mathbb{P}\{z=-1|\theta=+1\}=0.0007$
 &
 $\mathbb{P}\{z=-1|\theta=+1\}=0.3449$
 &
 $\mathbb{P}\{z=-1|\theta=+1\}=0.1526$
 \\
     \begin{tikzpicture}
     \node[] at (0,0) {\includegraphics[width=.6\columnwidth,trim={3cm 0.8cm 3cm 0.8cm},clip]{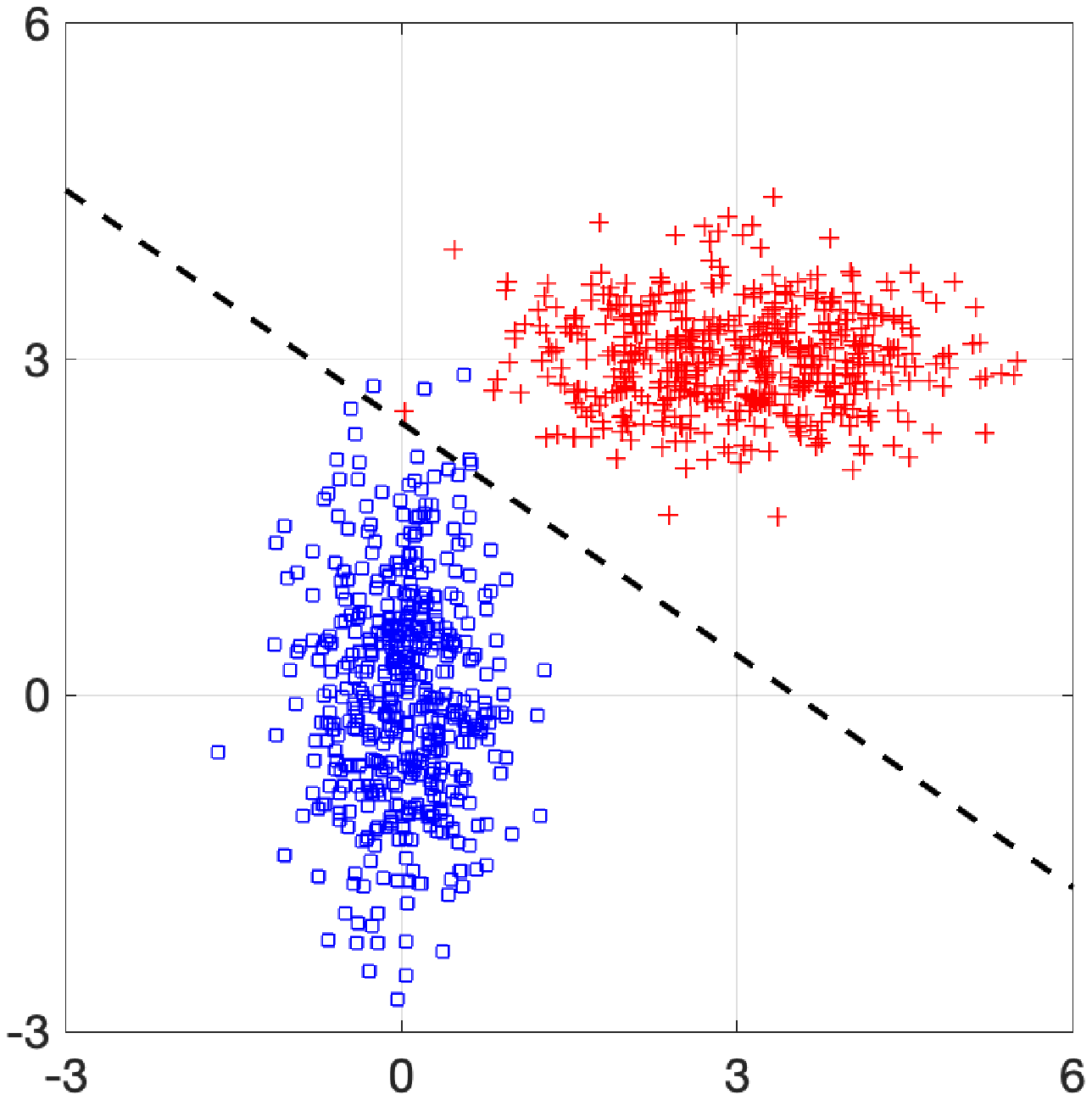}};
     \node[] at (0,-2.7) {$x_1$};
     \node[rotate=90] at (-2.7,0) {$x_2$};
     \end{tikzpicture}
     &
     \hspace{-.1in}
     \begin{tikzpicture}
     \node[] at (0,0) {\includegraphics[width=.6\columnwidth,trim={3cm 0.8cm 3cm 0.8cm},clip]{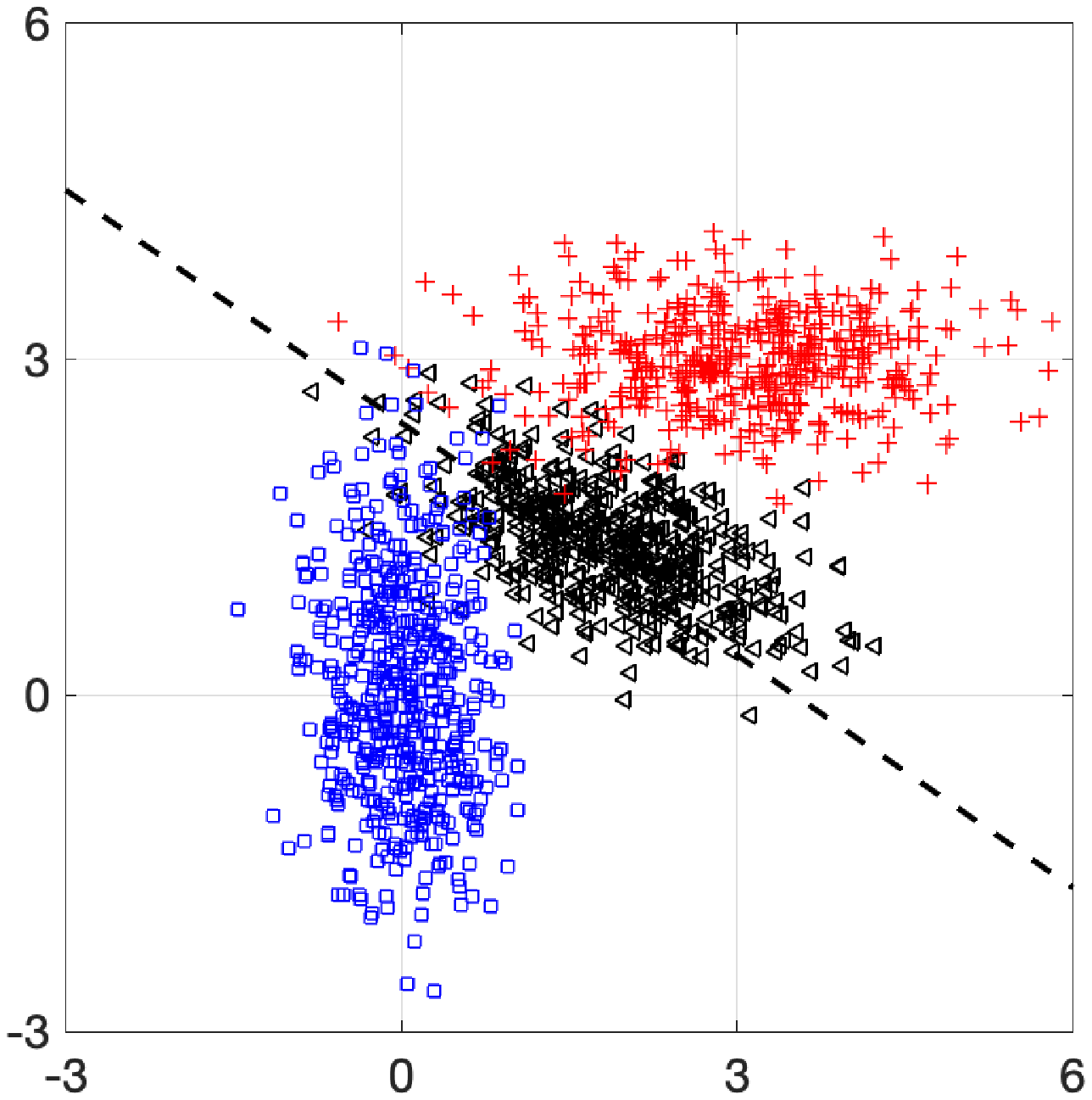}};
     \node[] at (0,-2.7) {$x_1$};
     \node[rotate=90] at (-2.7,0) {$x_2$};
     \end{tikzpicture}
     &
     \hspace{-.1in}
     \begin{tikzpicture}
     \node[] at (0,0) {\includegraphics[width=.6\columnwidth,trim={3cm 0.8cm 3cm 0.8cm},clip]{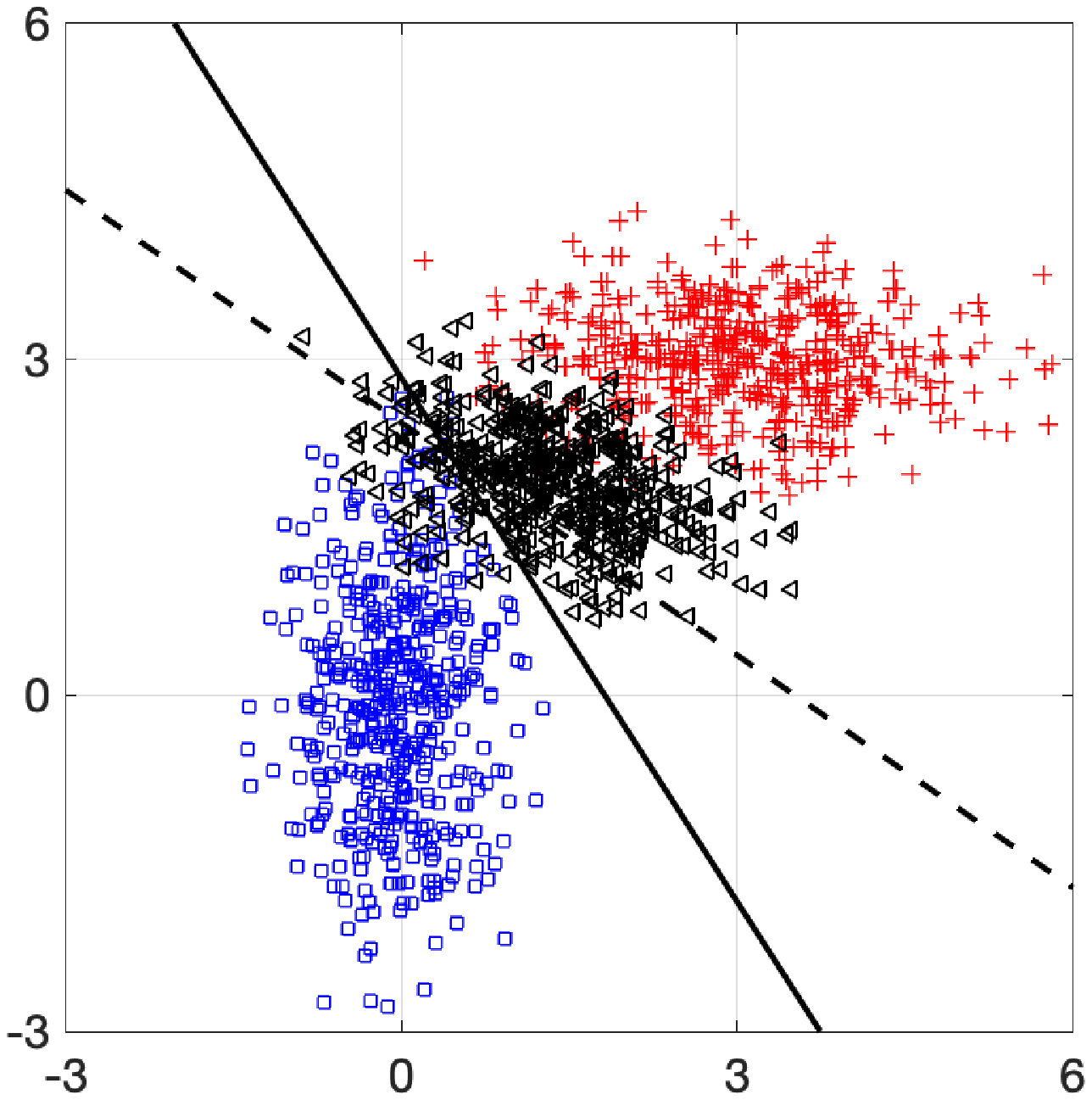}};
     \node[] at (0,-2.7) {$x_1$};
     \node[rotate=90] at (-2.7,0) {$x_2$};
     \end{tikzpicture}
     \\
     (a) Non-adversarial 
     &
     (b) Na\"{i}ve classifier 
     &
     (c) Equilibrium
 \end{tabular}
 \caption{
\label{fig:1} 
Illustration of the effect of the adversary on linear support vector machine classifiers using synthetic Gaussian data. There is no adversary in the non-adversarial case. The na\"{i}vete case refers to the case that the classifier is not prepared to respond to the adversary. The equilibrium of the game is recovered by Algorithm~\ref{alg:1}. The optimal non-adversarial and adversarial support vector machines are illustrated with dashed and solid lines, respectively. The original data points from the positive class are shown by the red pluses and the data points from the negative class are shown by the blue squares. The manipulated data points from the positive class are shown by the black triangles. }
\end{figure*}

\section{Numerical Example} \label{sec:numerical}
In this section, we illustrate the applicability of the developed game-theoretic framework on two numerical problems: an illustrative example using Gaussian data and a practical example using real data on heart disease classification. 
\subsection{Illustrative Example}
Consider an example in which the data for the positive and negative classes are Gaussian random variables with the following means and co-variance matrices:
\begin{align*}
\mu_+=
\begin{bmatrix}
3 \\
3
\end{bmatrix},
\mu_-=
\begin{bmatrix}
0 \\
0
\end{bmatrix},
\Sigma_+=
\begin{bmatrix}
1 & 0 \\
0 & 1/5
\end{bmatrix},
\Sigma_-=
\begin{bmatrix}
1/5 & 0 \\
0 & 1
\end{bmatrix}.
\end{align*}
For the following experiments 500 data points from each class are randomly generated to illustrate the effects of the adversary and the classifier. 

We first consider there is no adversary. In this case, the classifier is merely interested in identifying the optimal linear support vector machine by maximizing the true positive probability subject to a constraint that the true negative probability is greater than or equal to $1-\delta=0.99$. Figure~\ref{fig:1}~(a) shows the optimal non-adversarial support vector machine with the dashed line. The data points from the positive class are shown by the red pluses and the data points from the negative class are shown by the blue squares. In this case, the true negative probability is $0.99$ and the true positive probability (which is equal to one minus the false negative probability) is $0.9993$. This high performance is of course an artifact of the setup of the example in which the data points for both classes are linearly separable with high probability and mixing is only due to  outliers. 

Now, let the classifier select this optimal  non-adversarial support vector machine as its policy. Assume that the adversary can now manipulate the data in the positive class  but the classifier is not prepared to respond to these manipulations. The adversary is interested in manipulating the data from the positive class in order to maximize the false negative probability subject to a bound on the expected variance of the changes with $\epsilon=2$. Figure~\ref{fig:1}~(b) the effect of the adversary on the performance of this na\"{i}ve classifier. The na\"{i}vete refers to that the classifier is not prepared to respond to the adversary. The manipulated data points from the positive class are depicted by the black triangles. Similarly, the original data points from the positive class are shown by red pluses and the data points from the negative class are shown by the blue squares. In this case, the false negative probability increases 493\% in comparison to the non-adversarial case. This is of course due to unprepared, na\"{i}ve nature of the classifier. 

Now, consider the case where the classifier uses the optimal linear support vector machine extracted from the equilibrium of the game extracted by Algorithm~\ref{alg:1}. In this case, it is also in the benefit of the adversary to employ its optimal manipulation policy corresponding to the equilibrium of the game. Figure~\ref{fig:1}~(c) shows the optimal adversarial support vector machine with the solid line (compare with the non-adversarial support vector machine depicted still by the dashed line). Again, the manipulated data points from the positive class are shown by the black triangles, the original data points from the positive class are shown by red pluses, and the data points from the negative class are shown by the blue squares. By employing the optimal adversarial support vector machine, the classifier can reduce the false negative probability 56\% in comparison to the na\"{i}ve case.

\begin{table}
\centering
\caption{ \label{table:1}
The effect of adversary on linear support vector machine classifiers using the Cardiovascular Disease dataset. There is no adversary in the non-adversarial case. The na\"{i}vete case refers to the case that the classifier is not prepared to respond to the adversary. The equilibrium of the game is recovered by Algorithm~\ref{alg:1}. 
}
\begin{tabular}{c|c|c|}
\cline{2-3} 
 & True negative probability & False negative probability \\
  & $\mathbb{P}\{z=-1|\theta=-1\}$ &  $\mathbb{P}\{z=-1|\theta=+1\}$ 
 \\ \cline{1-3}
\multicolumn{1}{ |c| }{Non-adversarial} & 0.8986 & 0.2000 \\ \cline{1-3}
\multicolumn{1}{ |c| }{Na\"{i}ve classifier } & 0.8986 & 0.9758 \\ \cline{1-3}
\multicolumn{1}{ |c| }{Equilibrium} & 0.9130 & 0.6303 \\ \cline{1-3}
\end{tabular}
\end{table}

\subsection{Heart Disease}
In this section, we use the Cardiovascular Disease dataset on Kaggle~\cite{Cardiovascular}. The dataset contains age, height, weight, gender, Systolic and Diastolic blood pressures, Cholesterol level, Glucose level as well as smocking, drinking, and activity levels of 70,000 individuals.There are two classes of individuals: those with no heart disease (negative class) and those with a heart disease (positive class). 

Consider the case where an adversary is interested in miss-classifying an individual from the positive class. This could be motivated by that the adversary wants to forge medical documents with minimal changes to pass a life insurance test. For the sake of numerical stability, we scale the data with the inverse of the Cholesky factor of the co-variance matrix of the data. This is just to ensure that all the entries of the data are of the same size. This is indeed a linear transformation whose effect can be compensated for in the linear support vector machine and the linear manipulations of the adversary. Therefore, there is no loss of generality in using this transformation. 

To be able to use the developed framework, we fit Gaussian density functions to data points for each class and follow the approach of this paper for designing the classifier and computing the optimal manipulation by the adversary. In what follows, we set $\delta=10^{-1}$ and $\epsilon=1$. Table~\ref{table:1} shows the effect of the adversary on the linear support vector machine classifier. Although our Gaussian assumption might not be entirely valid, the optimal non-adversarial support vector machine almost meets the constraints on the true negative probability (with true negative probability of $0.8986$ instead of $1-\delta=0.9000$). In the na\"{i}ve case, the adversary can use the ignorance of the classifier to improve the false negative probability by 488\%; nearly all individuals from the positive class can be made to pass the test. Following the equilibrium extracted by Algorithm~\ref{alg:1}, the performance of the adversary is degraded by 35\%. This is of course a significant improvement for the classifier. To be able to further reduce the false negative probability, we need to increase $\delta$. This portrays the trade-off that the classifier faces.  

\section{Conclusions and Future Work} \label{sec:conclusions}
We used a constant-sum game to model the interaction between an adversary and a classifier. For Gaussian data and linear support vector machine classifiers, we transformed the optimization problems of the adversary and the classifier to convex optimization problems. We then utilized best response dynamics to learn an equilibrium of the game in order to extract linear support vector machine classifiers that are robust to adversarial input manipulations.

\bibliographystyle{ieeetr}
\bibliography{reference}

\begin{thebibliography}{10}

\bibitem{McKinseyreport}
{McKinsey Global Institute}, ``Notes from the {AI} frontier: Modeling the
  impact of {AI} on the world economy,'' 2018.

\bibitem{dalvi2004adversarial}
N.~Dalvi, P.~Domingos, S.~Sanghai, and D.~Verma, ``Adversarial
  classification,'' in {\em Proceedings of the tenth ACM SIGKDD International
  Conference on Knowledge Discovery and Data Mining}, pp.~99--108, ACM, 2004.

\bibitem{goodfellow2014explaining}
I.~J. Goodfellow, J.~Shlens, and C.~Szegedy, ``Explaining and harnessing
  adversarial examples,'' in {\em Proceedings of the 3rd International
  Conference on Learning Representations}, 2015.

\bibitem{yuan2019adversarial}
X.~Yuan, P.~He, Q.~Zhu, and X.~Li, ``Adversarial examples: Attacks and defenses
  for deep learning,'' {\em IEEE transactions on Neural Networks and Learning
  Systems}, 2019.

\bibitem{biggio2013evasion}
B.~Biggio, I.~Corona, D.~Maiorca, B.~Nelson, N.~{\v{S}}rndi{\'c}, P.~Laskov,
  G.~Giacinto, and F.~Roli, ``Evasion attacks against machine learning at test
  time,'' in {\em Joint European conference on machine learning and knowledge
  discovery in databases}, pp.~387--402, Springer, 2013.

\bibitem{papernot2016distillation}
N.~Papernot, P.~McDaniel, X.~Wu, S.~Jha, and A.~Swami, ``Distillation as a
  defense to adversarial perturbations against deep neural networks,'' in {\em
  2016 IEEE Symposium on Security and Privacy (SP)}, pp.~582--597, IEEE, 2016.

\bibitem{kurakin2016adversarial}
A.~Kurakin, I.~Goodfellow, and S.~Bengio, ``Adversarial machine learning at
  scale,'' in {\em Proceedings of the 5rd International Conference on Learning
  Representations}, 2017.

\bibitem{frederickson2018attack}
C.~Frederickson, M.~Moore, G.~Dawson, and R.~Polikar, ``Attack strength vs.
  detectability dilemma in adversarial machine learning,'' in {\em 2018
  International Joint Conference on Neural Networks (IJCNN)}, pp.~1--8, IEEE,
  2018.

\bibitem{han2018adequacy}
Y.~Han and B.~Rubinstein, ``Adequacy of the gradient-descent method for
  classifier evasion attacks,'' in {\em Workshops at the Thirty-Second AAAI
  Conference on Artificial Intelligence}, 2018.

\bibitem{russu2016secure}
P.~Russu, A.~Demontis, B.~Biggio, G.~Fumera, and F.~Roli, ``Secure kernel
  machines against evasion attacks,'' in {\em Proceedings of the 2016 ACM
  Workshop on Artificial Intelligence and Security}, pp.~59--69, ACM, 2016.

\bibitem{crawford1982strategic}
V.~P. Crawford and J.~Sobel, ``Strategic information transmission,'' {\em
  Econometrica: Journal of the Econometric Society}, pp.~1431--1451, 1982.

\bibitem{farokhi2016estimation}
F.~Farokhi, A.~M.~H. Teixeira, and C.~Langbort, ``Estimation with strategic
  sensors,'' {\em IEEE Transactions on Automatic Control}, vol.~62, no.~2,
  pp.~724--739, 2016.

\bibitem{farrell1996cheap}
J.~Farrell and M.~Rabin, ``Cheap talk,'' {\em Journal of Economic
  Perspectives}, vol.~10, no.~3, pp.~103--118, 1996.

\bibitem{kamenica2011bayesian}
E.~Kamenica and M.~Gentzkow, ``Bayesian persuasion,'' {\em American Economic
  Review}, vol.~101, no.~6, pp.~2590--2615, 2011.

\bibitem{dughmi2016algorithmic}
S.~Dughmi and H.~Xu, ``Algorithmic bayesian persuasion,'' in {\em Proceedings
  of the forty-eighth annual ACM symposium on Theory of Computing},
  pp.~412--425, ACM, 2016.

\bibitem{nadendla2018effects}
V.~S.~S. Nadendla, C.~Langbort, and T.~Ba{\c{s}}ar, ``Effects of subjective
  biases on strategic information transmission,'' {\em IEEE Transactions on
  Communications}, vol.~66, no.~12, pp.~6040--6049, 2018.

\bibitem{arrow1954existence}
K.~J. Arrow and G.~Debreu, ``Existence of an equilibrium for a competitive
  economy,'' {\em Econometrica: Journal of the Econometric Society},
  pp.~265--290, 1954.

\bibitem{schaible1974parameter}
S.~Schaible, ``Parameter-free convex equivalent and dual programs of fractional
  programming problems,'' {\em Zeitschrift f{\"u}r Operations Research},
  vol.~18, no.~5, pp.~187--196, 1974.

\bibitem{10230740590698}
E.~N. Barron, R.~Goebel, and R.~R. Jensen, ``Best response dynamics for
  continuous games,'' {\em Proceedings of the American Mathematical Society},
  vol.~138, no.~3, pp.~1069--1083, 2010.

\bibitem{Cardiovascular}
S.~Ulianova, ``Cardiovascular disease dataset: The dataset consists of 70 000
  records of patients data, 11 features + target.,'' 2019.
\newblock
  \url{https://www.kaggle.com/sulianova/cardiovascular-disease-dataset}.

\end{thebibliography}

\end{document}